\documentclass[prb,twocolumn,showpacs,preprintnumbers,amsmath,amssymb]{revtex4}
\usepackage{graphicx}% Include figure files
\usepackage{dcolumn}% Align table columns on decimal point
\usepackage{bm}% bold math

\newcommand{\mapright}[1]{\smash{\mathop{\hbox to 1.0cm{\rightarrowfill}}\limits^{#1}}}

\begin{document}

%\preprint{}

\
\title{Macroscopic quantum tunneling and quasiparticle-tunneling blockade effect in $s$-wave/$d$-wave hybrid junctions}% Force line breaks with \\

\author{S.~Kawabata,$^{1,2}$ A.~A.~Golubov,$^1$ Ariando,$^1$ C.~J.~M.~Verwijs,$^1$ H.~Hilgenkamp,$^1$ and J.~R.~Kirtley$^3$}
% \altaffiliation[Also at ]{Physics Department, XYZ University.}%Lines break automatically or can be forced with \\
%\author{Second Author}%
%\email{s-kawabata@aist.go.jp}
\affiliation{%
$^1$Faculty of Science and Technology, University of Twente, 
P.O. Box 217, 7500 AE Enschede, The Netherlands \\
$^2$Nanotechnology Research Institute (NRI), National Institute of 
Advanced Industrial Science and Technology (AIST), Tsukuba, 
Ibaraki, 305-8568, Japan \\
$^3$IBM T. J. Watson Research Center, Yorktown Heights, New York 10598, USA
}%

\date{\today}

\begin{abstract}
We have theoretically investigated macroscopic quantum tunneling (MQT) and the influence of nodal quasiparticles and zero energy bound states (ZES)
on MQT in  $s$-wave/ $d$-wave hybrid Josephson junctions.
In contrast to $d$-wave/$d$-wave junctions, the low-energy quasiparticle dissipation resulting from nodal quasiparticles and ZESs is suppressed due to a quasiparticle-tunneling blockade effect in an isotropic $s$-wave superconductor. 
Therefore, the inherent dissipation in these junctions is found to be weak. 
We have also investigated MQT in a realistic $s$-wave/$d$-wave (Nb/Au/YBCO) junction in which Ohmic dissipation in a shunt resistance is stronger than the inherent dissipation and find that MQT is observable within the current experimental technology.
This result suggests high potential of $s$-wave/$d$-wave hybrid junctions for applications in quantum information devices.
\end{abstract}

% PACS, the Physics and Astronomy
                             % Classification Scheme.
%\keywords{Suggested keywords}%Use showkeys class option if keyword
                              %display desired
\maketitle

%%%
%s-wave/d-wave->sd
%
%

\section{Introduction}

Since the experimental observations of macroscopic quantum tunneling (MQT) in YBCO (YBaCuO) grain boundary~\cite{rf:Bauch1,rf:Bauch2} and BSCCO (BiSrCaCuO) intrinsic~\cite{rf:Inomata,rf:Jin,rf:KashiwayaMQT1,rf:Inomata2,rf:KashiwayaMQT2} Josephson junctions, the macroscopic quantum dynamics of high-$T_c$ $d$-wave junctions has become a hot topic in the field of superconductor quantum electronics and quantum computation.  
Recently, the effect of low-energy quasiparticles, e.g., the nodal-quasiparticles and the zero energy bound states (ZESs)~\cite{rf:Kashiwaya,rf:Lofwander} on MQT in $d$-wave  junctions have been theoretically investigated.~\cite{rf:Kawabata1,rf:Yokoyama,rf:Kawabata2,rf:Kawabata3,rf:Kawabata4}
It was found that, in $c$-axis type junctions,~\cite{rf:Yurgens} the suppression of MQT resulting from the nodal-quasiparticles is very weak.~\cite{rf:Kawabata1,rf:Yokoyama} 
This result is consistent with recent experimental observations.~\cite{rf:Inomata,rf:Jin,rf:KashiwayaMQT1,rf:Inomata2,rf:KashiwayaMQT2} 
In the case of in-plane type $d$-wave junctions,~\cite{rf:Hilgenkamp} however, the ZESs give a strong dissipative effect.~\cite{rf:Kawabata2,rf:Kawabata3,rf:Kawabata4} 
Therefore, it is important to avoid the formation of ZESs in order to observe MQT with a high quantum-to-classical crossover temperature $T^*$.

On the other hand, recently, quiet qubits consisting of a superconducting loop with an $s$-wave/$d$-wave ($s/d$) hybrid junction have been proposed.~\cite{rf:Ioffe,rf:Blatter,rf:Zagoskin}
In quiet qubits, a quantum two level system is spontaneously generated and therefore it is expected to be robust to decoherence from fluctuations of the external magnetic field.~\cite{rf:Ioffe,rf:Blatter,rf:Kawabata5}
However, the influence of the low-energy quasiparticle dissipation due to nodal-quasiparticles and ZESs on $s/d$ quiet qubits is not yet understood.
Therefore, it is important to investigate such intrinsic dissipation effects on the macroscopic quantum dynamics in order to realize quiet qubits.

 In this paper, motivated by the above studies and the recent phase-sensitive spectroscopy experiments in $s/d$ junctions,~\cite{rf:Smilde,rf:Kirtley} we will discuss the application of a generic MQT theory~\cite{rf:Esteve,rf:Zaikin} to $s/d$ hybrid Josephson junctions [see Fig. 1(a)].
  In contrast with $d/d$ junctions,~\cite{rf:Kawabata1,rf:Kawabata2,rf:Kawabata3} we will show that the influence of the low-energy quasiparticle dissipation is suppressed by virtue of the small but finite isotropic gap in the $s$-wave superconductor. 
 Therefore, weak quasiparticle dissipation is anticipated in such junctions.
We will also discuss an extrinsic Ohmic dissipative effect in a real $s/d$ (Nb/Au/YBCO) junction~\cite{rf:Smilde,rf:Kirtley} and show that high $T^*$,  comparable to high-quality $s/s$ junctions, is expected.
  These results clearly indicate the advantage of $s/d$ junction for qubit applications with longer coherence time.

\section{Theory of macroscopic quantum tunneling}
\subsection{Macroscopic quantum tunneling in $s$-wave/$d$-wave junctions}
In the following we will calculate the MQT rate in $s$/$d$ junctions with a clean insulating barrier and without extrinsic dissipation, e.g., an Ohmic dissipation in a shunt resistance.
Extrinsic dissipative effects in an actual junction will be discussed later.
The partition function of a junction can be described by a functional integral over the macroscopic variable (the phase difference $\phi$),~\cite{rf:Zaikin,rf:Ambegaokar,rf:Eckern}  i.e., 
\begin{eqnarray}
{\cal  Z} 
= 
\int 
{\cal D} \phi (\tau) 
\exp
\left(
  - \frac{{\cal S}_{\mathrm{eff}}[\phi]}{\hbar}
\right)
.
\end{eqnarray}
%
%
%
%
 %%%
%%%
%%%
%%%
%===================================
\begin{figure}[t]
\begin{center}
\includegraphics[width=8.7cm]{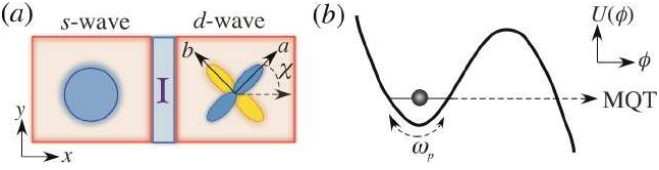}
\caption{(Color online) (a) Schematic of the in-plane $s$-wave/$d$-wave hybrid Josephson junction. $a$ and $b$ denote the crystalline axes of the $d$-wave superconductor, and $\chi$ is the mismatch angle between the normal of the insulating barrier (I) and the $a$-axis.
(b) Potential $U(\phi)$ v.s. the phase difference  $\phi$ between two superconductors.
$\omega_p$ is the Josephson plasma frequency of the junction.
}
\end{center}
\end{figure}
%===================================
%%%
%%%
%%%
%%%
In the high barrier limit, i.e., $z_0\equiv m w_0/\hbar^2 k_F  \gg 1$ ($m$ is the electron mass, $w_0$ is the height of the delta function-type insulating-barrier I (see Fig. 1(a)), and $k_F$ is the Fermi wave length), the effective action ${\cal S}_{\mathrm{eff}}$ is given by 
\begin{eqnarray}
{\cal S}_{\mathrm{eff}}[\phi]
&= &
\int_{0}^{\hbar \beta} d \tau 
\left[
   \frac{M}{2} 
   \left(
   \frac{\partial \phi ( \tau) }{\partial \tau}
   \right)^2
   + 
   U(\phi)
\right] + S_\alpha[\phi],
\nonumber\\
{\cal S}_\alpha[\phi]
&=&
-
\int_{0}^{\hbar \beta} d \tau  \int_{0}^{\hbar \beta} d \tau'
  \alpha (\tau - \tau') \cos \frac{\phi(\tau) - \phi (\tau') }{2}
.
\nonumber\\
  \label{eqn:alpha2}
\end{eqnarray}
In this equation, $\beta = 1 /k_B T$, $M
 = 
 C \left( \hbar/2 e\right)^2
$
is the mass ($C$ is the capacitance of the junction) and the potential $U(\phi)$ can be described by 
\begin{eqnarray}
 U(\phi) 
 = 
\frac{\hbar }{2e} 
\left[
    \int_0^1 d \lambda \ \phi  I_J (\lambda \phi) - \phi \  I_{ext}
\right],
\end{eqnarray}
where $I_J$ is the Josephson current and $ I_{ext}$ is the external bias current, respectively.
The dissipation kernel $\alpha(\tau)$ is related to the quasiparticle current $I_{\mathrm{qp}}$
under constant bias voltage $V$ by
\begin{eqnarray}
\alpha(\tau) 
=
\frac{\hbar }{e}
\int_0^\infty \frac{d \omega}{2 \pi}  \exp \left(  -\omega \tau \right)
    I_{\mathrm{qp}} \left( V=\frac{\hbar \omega}{e} \right)
,
\end{eqnarray}
at zero temperature.~\cite{rf:Zaikin,rf:Ambegaokar,rf:Eckern}

Below, in order to investigate the effect of the nodal-quasiparticles and the ZESs on MQT, we derive the effective action for two types of $s$/$d_\chi$ junctions, i.e., $\chi=0$ and $\pi/4$, with $\chi$ being the mismatch angle between the normal of the insulating barrier (I) and the crystalline axis of the $d$-wave superconductor (see Fig. 1(a)).
In the case of $s/d_0$ junctions, ZESs are completely absent.~\cite{rf:Kashiwaya}
On the other hand, in $s/d_{\pi/4}$ junctions, ZES are formed near the interface between I and the $d$-wave superconductor $d_{\pi/4}$.
Note that the influence of the ZES is maximized at $\chi=\pi/4$ in the range $0 \le \chi \le \pi/2$.~\cite{rf:Kashiwaya}

 First, we will calculate the potential energy $U$ which can be described by the Josephson current flowing through the junction (Eq. (3)).
 In order to obtain the Josephson current, we solve the Bogoliubov-de Gennes equation with appropriate boundary conditions.~\cite{rf:Kashiwaya,rf:Lofwander}
Then, in the case of low temperatures ($\beta^{-1}\ll \Delta_0$) and high barriers ($z_0 \gg 1$), we get the analytical expressions as
\begin{eqnarray}
 I_J(\phi) 
\approx 
 \left\{
\begin{array}{cl}
\displaystyle{I_{C_1} \sin \phi}
&
\mbox{for} \ \   \mbox{$s/d_0$}
 \\
\displaystyle{- I_{C_2} \sin 2 \phi}
&    
\mbox{for} \ \   \mbox{$s/d_{\pi/4}$}
\end{array}
\right.
,
\end{eqnarray}
 where 
\begin{eqnarray}
 I_{C_1} &=& \frac{3  }{2 \pi}  I_0  \varepsilon  \int_{-1}^1 d x \frac{(1+x)x}{\sqrt{1-x} }  K(\sqrt{1- x^2 \varepsilon^2}),
\\
I_{C_2} &= & \frac{3}{10} \frac{ \Delta_d \beta I_0 R_Q  }{   N_c R_N} \varepsilon ,
\end{eqnarray}
$\varepsilon \equiv \Delta_d/\Delta_s$, $K$ is the complete elliptic integral of the second kind, $I_0=\pi \Delta_s/2 e R_N$ is the Josephson critical current for the $s$/$s$ junctions, $R_N$ is the normal state resistance of the junction, $R_Q=h/4 e^2$ is the resistance quantum,  and $N_c $ is the channel number at the Fermi energy.
In this calculation, we have assumed that the amplitude of the pair potential is given by $\Delta_s$ for $s$, $\Delta_d \cos 2 \theta \equiv \Delta_{d_0}(\theta)$ for $d_0$, and $\Delta_d \sin 2 \theta \equiv \Delta_{d_{\pi/4}}(\theta)$ for $d_{\pi/4}$, where $\cos \theta = k_x/k_F$.
By using the analytical expression of the Josephson current (5), we get
\begin{eqnarray}
U(\phi) 
\approx 
\left\{
\begin{array}{cl}
\displaystyle{- \frac{\hbar I_{C_1}}{2e}\left(  \cos \phi + \eta \phi \right) }
&    
\mbox{for} \ \   \mbox{$s/d_0$}
 \\
\displaystyle{- \frac{\hbar I_{C_2}}{4e}\left( -  \cos   2 \phi  + 2 \eta \phi  \right)}    
&    
\mbox{for} \ \   \mbox{$s/d_{\pi/4}$}
\end{array}
\right.
,
\end{eqnarray}
where $\eta \equiv I_{\mathrm{ext}}/I_{C_1(C_2)}$.
As in the case of $s$/$s$~\cite{rf:Zaikin,rf:Ambegaokar,rf:Eckern}  and $d$/$d$~\cite{rf:Kawabata1,rf:Kawabata2} junctions, $U$ can be expressed as a tilted washboard potential (see Fig.~1(b)).

Next we will calculate the dissipation kernel $\alpha(\tau)$.
In the high barrier limit, the quasiparticle current is given in terms of the convolution of the quasiparticle surface density of states (DOS),~\cite{rf:Kashiwaya,rf:Lofwander} i.e., 
\begin{eqnarray}
I_{\mathrm{qp}}(V)
&=& \frac{2e}{h} \sum_{k_y} |t_I|^2 \int_{-\infty}^{\infty}dE N_{L} (E,\theta)  N_{R} (E+ eV, \theta) 
\nonumber\\
&\times&
\left[
 f(E) -  f(E +eV)
\right]
,
\end{eqnarray}
where $t_I \approx \cos \theta /z_0$ is the transmission amplitude of the barrier I, $N_{L(R)} (E,\theta)$ is the quasiparticle surface DOS ($L=s$ and $R=d_0$ or $d_{\pi/4}$), and $f(E)$ is the Fermi-Dirac distribution function.
The surface DOS for $s$-wave superconductors is
\begin{eqnarray}
 N_s(E,\theta)= N_s(E)= \mbox{Re} \left( \frac{ |E|}{\sqrt{E^2-\Delta_{s}^2}} \right).
 \end{eqnarray}
In the case of $d_0$, no ZES are formed.
Therefore the angle $\theta$ dependence of the DOS is the same as the bulk, i.e., 
\begin{eqnarray}
 N_{d_{0}} (E,\theta)
=     \mbox{Re} \left( \frac{|E| }{ \sqrt{E^2-\Delta_{d_0}(\theta)^2}} \right )
.
\end{eqnarray}
On the other hand, the DOS for $d_{\pi/4}$ is given by~\cite{rf:Matsumoto}
\begin{eqnarray}
 N_{d_{\pi/4}} (E,\theta)
& =&     \mbox{Re} \left(   \frac{\sqrt{E^2 - \Delta_{d_{\pi/4} }(\theta)^2 }}{|E|}  \right)
  \nonumber\\
& +&      \pi | \Delta_{d_{\pi/4}}(\theta)| \delta(E)
 .
\end{eqnarray}
The delta function peak at $E=0$ corresponds to the ZES. 
Because of the quasiparticle bound state at $E=0$, the quasiparticle current for the $s/d_{\pi/4}$ junctions is drastically different from that for the $s/d_0$ junctions in which no ZES are formed.~\cite{rf:Lofwander}
In the limit of low temperatures ($\beta^{-1} \ll \Delta_s$), we can obtain the analytical expression of the dissipation kernel $\alpha(\tau)$ as
\begin{eqnarray}
\alpha(\tau)
\approx 
\left\{
\begin{array}{cl}
\displaystyle{
  \frac{3 \hbar}{8 \sqrt{2} \pi}
\frac{ R_Q }{ R_N \varepsilon}
\frac{1}{\tau^2}
K_1 \left(  \frac{\Delta_s |\tau|}{\hbar} \right)
   }
& 
\    \mbox{for} \quad \mbox{$s/d_{0}$}\\
\displaystyle{
  \frac{6}{5 \hbar}
\frac{ R_Q \Delta_s \Delta_d}{ R_N }
K_1 \left(  \frac{\Delta_s |\tau|}{\hbar} \right)   }
& 
\   \mbox{for}\quad   \mbox{$s/d_{\pi/4}$}\\
\end{array}
\right.
,
\end{eqnarray}
where $K_1$ is the modified Bessel function.
For $ |\tau| \gg \hbar / \Delta_s$ the dissipation kernel decays exponentially as a function of the imaginary time $\tau$, i.e., 
\begin{eqnarray}
\alpha(\tau)
\approx 
\left\{
\begin{array}{c}
\displaystyle{
  \frac{3 \hbar^{3/2}}{16 \sqrt{\pi}}
\frac{ R_Q  \sqrt{\Delta_s}}{ R_N \Delta_d}
\frac{1}{\left| \tau \right|^{5/2}}
\exp \left( - \frac{\Delta_s |\tau|}{\hbar} \right)
   }
\\
\displaystyle{\quad \quad \quad   \quad \quad \quad   \mbox{for} \quad \mbox{$s/d_{0}$}}
\\
\displaystyle{
  \frac{6 \sqrt{\pi} }{5 \sqrt{2 \hbar}}
\frac{ R_Q \sqrt{\Delta_s} \Delta_d}{ R_N }
\frac{1}{\sqrt{\left| \tau \right|}}
\exp \left( - \frac{\Delta_s |\tau|}{\hbar} \right)
}
\\ 
\displaystyle{\quad \quad \quad   \quad \quad \quad   \mbox{for}\quad   \mbox{$s/d_{\pi/4}$}}
\\
\end{array}
\right.
.
\end{eqnarray}
The typical dynamical time scale of the macroscopic variable $\phi$ is of the order of the inverse Josephson plasma frequency $\omega_p$ which is much smaller than $\Delta_s$.
Thus the phase varies slowly with the time scale given by $\hbar /\Delta_s$, then we can expand $\phi(\tau) - \phi (\tau')$ in Eq. (\ref{eqn:alpha2}) about $\tau=\tau'$.
This gives 
\begin{eqnarray}
S_\alpha[\phi]
\approx
\frac{\delta C}{2}
\int_{0}^{\hbar \beta} d \tau 
   \left[
   \frac{\hbar}{2e}
   \frac{   \partial \phi ( \tau) }{\partial \tau}
   \right]^2
   .
\end{eqnarray}
Hence, the dissipation action ${\cal S}_\alpha$ acts as a kinetic term so that the effect of the quasiparticles results in an increase of the capacitance, $C \to C + \delta C \equiv C_{ren}$.
This indicates that the quasiparticle dissipation in $s$/$d$ junctions is qualitatively weaker than that in in-plane $d/d$ junctions in which the super-Ohmic ($\alpha(\tau) \sim |\tau|^{-3}$)\cite{rf:Kawabata1,rf:Yokoyama,rf:Bruder,rf:Barash,rf:Fominov,rf:Joglekar} or Ohmic dissipation ($\alpha(\tau) \sim \tau^{-2}$)~\cite{rf:Kawabata2,rf:Kawabata3,rf:Kawabata4,rf:Amin} appears.
At zero temperature, the capacitance increment $\delta C$ can be calculated as
\begin{eqnarray}
\delta C
=
\left\{
\begin{array}{cl}
\displaystyle{
  \frac{3 }{8 \sqrt{2} \pi}
\frac{ e^2 R_Q }{ \Delta_s R_N}
A(\varepsilon)   }
& 
\    \mbox{for} \quad \mbox{$s/d_{0}$}\\
\displaystyle{
  \frac{24}{5}
\frac{ e^2 R_Q}{  \Delta_s R_N }
\varepsilon}
& 
\   \mbox{for}\quad   \mbox{$s/d_{\pi/4}$}\\
\end{array}
\right.
.
\end{eqnarray}
In this equation 
\begin{eqnarray}
A(\varepsilon)&=&\int_{-1}^1 d x
 \frac{(1+x)(1- \varepsilon^2 x^2)^2}
 {\sqrt{1-x}}
  \left[
    (1+\varepsilon^2 x^2) 
    \right.
    \nonumber\\
    &\times&
    \left.
     E (\sqrt{1- \varepsilon^2 x^2}) 
    - 
    2 \varepsilon^2 x^2 K( \sqrt{1- \varepsilon^2 x^2})
    \right]
,
\end{eqnarray}
where $E$ is the complete elliptic integrals of the first kind.
Fig. 2 shows the dependence of the capacitance increase $\delta C$ on $\Delta_s$, where we have used $R_N=3.68\Omega$~\cite{rf:Smilde} and $\Delta_d=\Delta_\mathrm{YBCO} = 20.0$meV.
$\delta C$ is rapidly decreasing with increasing $\Delta_s$.
Thus, in the case of large $\Delta_s$, the influence of the low-energy quasiparticle dissipation on the macroscopic quantum dynamics becomes very weak.
This is a clear indication of quasiparticle-tunneling blockade effects in isotropic $s$-wave superconductors.
As an example, if $s=$Nb and $d=$ YBCO ($\Delta_s = 1.55$meV, and $\varepsilon^{-1}=\Delta_s /\Delta_d =0.0775$), we get $\delta C=9.5$fF for $s/d_0$ and $\delta C=11$pF for $s/d_{\pi/4}$ .
Therefore, in this case, the effect of the ZESs on the macroscopic quantum dynamics is considerably stronger than that of the nodal-quasiparticles.

%%%
%%%
%%%
%%%
%===================================
\begin{figure}[tb]
\begin{center}
\includegraphics[width=9.3cm]{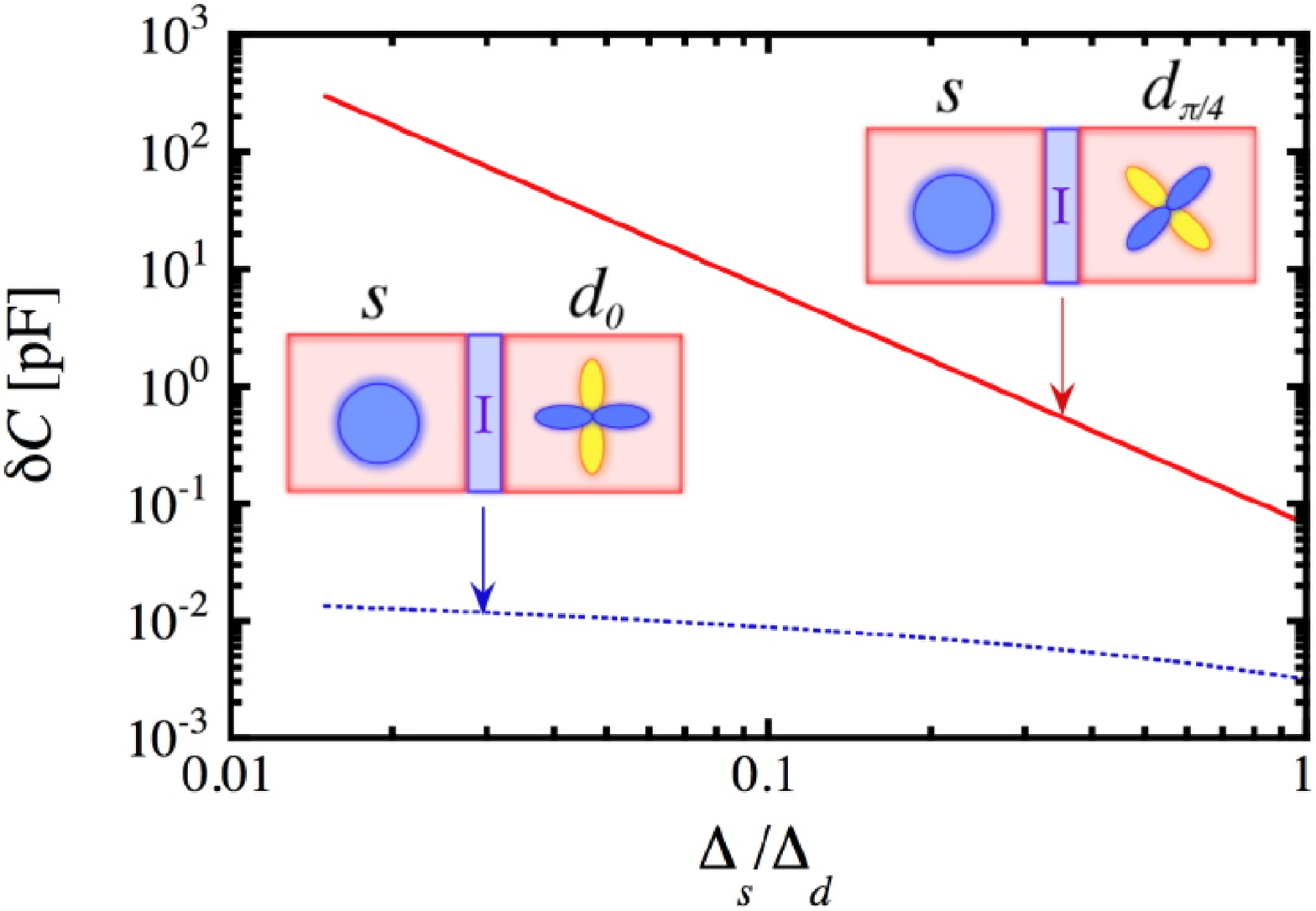}
\caption{(Color online) The $s$-wave gap $\Delta_s$ dependence of the capacitance increment $\delta C$ due to the low-energy quasiparticle dissipation for  $s$/$d_0$ (dashed line) and $s$/$d_{\pi/4}$ (solid line) junctions.
}
\end{center}
\end{figure}
%===================================
%%%
%%%
%%%
%%%

%
%
%
%
%\section{MQT and crossover temperature}
%
%
%
%
Next, we will investigate MQT in $s$/$d$ junctions
The MQT escape rate from the metastable potential (Fig. 1(b)) at zero temperature is given by~\cite{rf:MQT1}
\begin{eqnarray}
\Gamma
=
\lim_{\beta \to \infty} \frac{2}{\beta} \mbox{ Im}\ln {\cal Z}.
\end{eqnarray}
By using the Caldeira and Leggett theory,~\cite{rf:Caldeira} the MQT rate is approximated as 
\begin{eqnarray}
\Gamma(\eta)
=
\frac{\omega_p(\eta)}{2 \pi}\sqrt{120 \pi B(\eta) }\  \exp [ -B(\eta)]
,
\end{eqnarray}
where 
\begin{eqnarray}
\omega_p (\eta)=  \sqrt{\frac{\hbar a_i I_{C_i} }{2 e M_\mathrm{ren}}}(1-\eta^2)^{\frac{1}{4}},
\end{eqnarray}
is the Josephson plasma frequency ($i=1$ for $d_0$, $i=2$ for $d_{\pi/4}$, $a_1=1$, $a_2=2$, and $M_\mathrm{ren}=(\hbar/2 e)^2 C_\mathrm{ren}$) and $B(\eta)= {\cal S}_{\mathrm{eff}}[\phi_B]/\hbar$ is the bounce exponent, that is the value of the the action ${\cal S}_{\mathrm{eff}}$ evaluated along the bounce trajectory $\phi_B(\tau)$.
The analytic expression for the bounce exponent is given by
\begin{eqnarray}
B(\eta)=\frac{b_i}{e}
  \sqrt{ \frac{2 e}{\hbar}I_{C_i} M_\mathrm{ren}}
  \left(
  1 -
 \eta^2
\right)^{\frac{5}{4}}
,
\end{eqnarray}
where $b_1=12/5$ and $b_2=3\sqrt{2}/5$.
In actual MQT experiments, the switching current distribution $P(\eta)$ is measured. 
$P(\eta)$ is related to the MQT rate $\Gamma(\eta)$ as 
\begin{eqnarray}
 P(\eta)=\frac{1}{v}
\Gamma(\eta) \exp
\left[
-\frac{1}{v}
\int_0^{\eta} \Gamma(\eta') d \eta'
\right]
,
\end{eqnarray}
where $v \equiv \left| d \eta / d t \right| $ is the sweep rate of the external bias current.
At high temperatures, the thermally activated (TA) decay dominates the escape process.
Then the escape rate is given by the Kramers formula~\cite{rf:MQT1}
\begin{eqnarray}
\Gamma=\frac{\omega_p}{2 \pi }\exp \left( -\frac{ U_0}{ k_B T}\right),
\end{eqnarray}
where $U_0$ is the barrier height.
Below the crossover temperature $T^*$, the escape process is dominated by MQT. 
In the absence of dissipative effects, $T^* \propto \sqrt{j_C/(C/A)}$,~\cite{rf:Wallraff} where $j_C=I_c/A$ is the Josephson critical current density ($A$ is the junction area) and $C/A=\epsilon_I /d_I$ ($\epsilon_I$ is the permittivity of I and $d_I$ is the barrier thickness).
Importantly, $T^*$ is reduced in the presence of dissipation.~\cite{rf:MQT1,rf:Caldeira}

In order to see explicitly the effect of the quasiparticle dissipation on MQT, we numerically estimate $T^*$.
We determine  $T^*$ from the relation 
\begin{eqnarray}
\sigma^\mathrm{TA} (T^*) = \sigma^\mathrm{MQT},
\end{eqnarray}
where $\sigma^\mathrm{TA} (T)$ and $\sigma^\mathrm{MQT} $ are the standard deviation of $P(\eta)$ for the temperature-dependent TA process and  the temperature-independent MQT process, respectively.
Presently, no experimental data are available for ideal highly under-damped $s/d$ junctions with large McCumber parameters $\beta_M = (2e/\hbar) I_C C R_{\mathrm{sg}}^2 \gg 1$ ($R_{\mathrm{sg}}$ is the subgap resistance).
Thus, we estimate $T^*$ by using the parameters for an actual Nb/Au/YBCO junction~\cite{rf:Smilde} ( $C=0.60$pF, $ I_C= 95.2 \mu$A at $\chi=0$, and $R_N=3.68 \Omega$) in which $\beta_M \approx 1.5$.
We also assume $ I_{C_1}=I_{C_2}=95.2 \mu$A for simplicity, $\Delta_s=\Delta_\mathrm{Nb}=1.55$meV, $\Delta_d=\Delta_\mathrm{YBCO}=20.0$meV,~\cite{suppression} and $ v  I_{C_i} = 0.0424$A/s.
In the case of $s/d_0$ junctions, we obtain $T^* = 336 $mK for the dissipationless case ($C_\mathrm{ren}= C$) and $T^* = 333 $mK for the dissipative case ($C_\mathrm{ren}= C + \delta C$).
Thus, the influence of the nodal-quasiparticle is negligibly small.
On the other hand, in the case of $s/d_{\pi/4}$ junctions, we get $T^* = 601 $mK for the dissipationless case and $T^* = 101 $mK for the dissipative case.
Therefore, as expected, the ZES have a larger influence on MQT than the nodal-quasiparticles.
However, the $T^*$ suppression is small enough to allow experimental observations of MQT.

\subsection{Macroscopic quantum tunneling in Nb/Au/YBCO junctions}

In the above calculation, we have assumed that the junction is ideal, i.e., the insulating barrier (I) is perfect so that we have ignored dissipation by a shunt resistance which is caused by imperfections of I. 
However, in practice, it is currently very difficult to fabricate such a perfect  junction.
Here we will investigate MQT in realistic junctions to check the feasibility of the MQT observation.

Recently, Nb/Au/YBCO ramp-edge junctions have been fabricated using the pulsed laser deposition technique.~\cite{rf:Smilde,rf:SmildeAPL}
These junctions can be regarded as hybrid $s/d$ junctions between the proximity-effect induced $s$-wave superconductor in Au and the $d$-wave superconductor in YBCO.~\cite{rf:Smilde}
Importantly, the mismatch angle $\chi$ of the YBCO crystal can be artificially varied as a single parameter.
By use of these junctions, angle-resolved electron tunneling experiments~\cite{rf:Smilde} and scanning-SQUID microscopy measurements~\cite{rf:Kirtley} have been performed in order to test the momentum dependence of the Cooper pairing amplitude and phases in YBCO.

The main differences between these junctions and the junctions with the clean interface discussed in Sec. II A are as follows.
Firstly, in actual Nb/Au/YBCO junctions, (i) the ohmic dissipation in the shunt resistance will dominate over the intrinsic quasiparticle dissipation resulting from the nodal-quasiparticles and the ZESs, because the observed McCumber parameter $\beta_M$ is the order of unity.
This observation can not be explained only by taking into account an intrinsic quasiparticle dissipation.
Thus we have to consider an extrinsic Ohmic dissipation source.
Moreover, due to the surface roughness~\cite{rf:Kashiwaya,suppression} and the very low transparency~\cite{rf:Riedel} of the actual insulating barrier I, (ii) the contribution of the ZESs to the Josephson current is expected to be small, so the $\sin 2 \phi$ component of the Josephson current can be neglected.
From (ii), the actual $\chi$ dependence of the Josephson current is given by 
\begin{eqnarray}
I_J (\chi, \phi)=I_C (\chi) \sin \phi
,
\end{eqnarray}
with $ I_C (\chi \approx \pi/4) \approx 0$.~\cite{rf:Smilde}  
This  can be nicely fitted by a theoretical model (Eq. (1) in Ref. [\onlinecite{rf:Smilde}]) in which the effect of the ZESs is neglected.

Then the effective action which describes the Nb/Au/YBCO junction can be expressed as
\begin{eqnarray}
&&{\cal S}_{\mathrm{eff}}[\phi]
=
\int_{0}^{\hbar \beta} d \tau 
\left[
   \frac{M}{2} 
   \left(
  \frac{\partial \phi ( \tau) }{ \partial \tau}
  \right)^2
   +
   U(\phi) 
\right] 
\nonumber\\
&-&
\int_{0}^{\hbar \beta} d \tau  
\int_{0}^{\hbar \beta} d \tau'
  \alpha_{\mathrm{s}} (\tau - \tau') \frac{ \left[ \phi(\tau) - \phi (\tau') \right]^2}{8}
  ,
\end{eqnarray}
with 
\begin{eqnarray}
U(\phi) =- \frac{\hbar I_C (\chi) }{2 e}  \left(\cos \phi  + \eta \phi \right)
,
\end{eqnarray}
 where $\eta = $ $I_\mathrm{ext} / I_C (\chi)$.
The kernel describing the Ohmic dissipation in the shunt resistance can be modeled as~\cite{rf:Zaikin}  
\begin{eqnarray}
 \alpha_{\mathrm{s}}(\tau)= 
 \frac{ \hbar  }{\pi^2 }
 \frac{ R_Q }{R_\mathrm{s}}
  \frac{1}{ \tau^{2}}
 ,
 \end{eqnarray}
where the shunt resistance $R_\mathrm{s}$ typically corresponds to $R_{\mathrm{sg}}$.
Along the same lines of the method used above, one finds that the MQT rate for the Nb/Au/YBCO junction is given by
\begin{eqnarray}
\Gamma(\chi,\eta)
&=&
\frac{\omega_p(\chi,\eta)}{2 \pi} \sqrt{120 \pi B(\chi, \eta) } 
\nonumber\\ 
&\times& \exp \left[ -B(\chi,\eta)
+
\frac{54 \zeta(3) R_Q}{\pi^4 R_\mathrm{s}} (1-\eta^2) \right]
,
\label{eqn:MQTybco}
\end{eqnarray}
 where the plasma frequency and the bounce exponent is respectively given by 
\begin{eqnarray}
\omega_p(\chi,\eta)&=&\sqrt{ \frac{\hbar I_C (\chi)}{2 e M}}  (1-\eta^2)^{\frac{1}{4}}
, \\
B(\chi, \eta)&=& \frac{12}{5e} \sqrt{ \frac{2e}{\hbar} I_C(\chi) M} (1-\eta^2)^{\frac{5}{4}},
\end{eqnarray}
and $\zeta(3)$ is the Riemann zeta function.
The second term in the exponent results from Ohmic dissipation in the shunt resistance.

%%%
%%%
%%%
%%%
%===================================
\begin{figure}[t]
\begin{center}
\includegraphics[width=9.2cm]{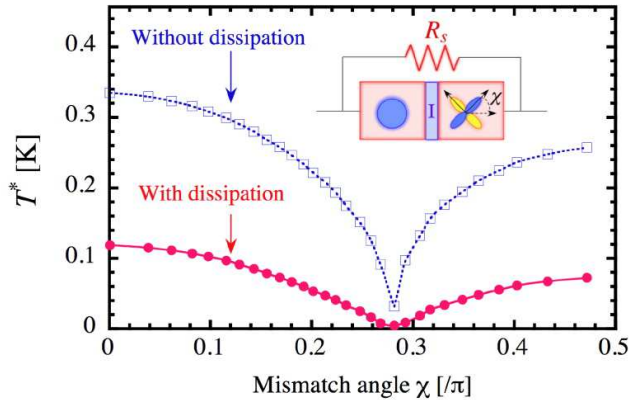}
\caption{(Color online) The mismatch angle $\chi$ dependence of the crossover temperature $T^*$ for an actual Nb/Au/YBCO junction with (closed circles) and without (open squares) the Ohmic dissipation in a shunt resistance $R_s$ (inset). 
}
\end{center}
\end{figure}
%===================================
%%%
%%%
%%%
%%%

From the MQT rate formula [Eq. (\ref{eqn:MQTybco})], we numerically estimate the $\chi$ dependence of $T^*$.
We have extracted $I_C (\chi)$ from the theoretical fitting curve of the experimental data (see Fig. 3(d) in Ref. [\onlinecite{rf:Smilde}]) and used $C=0.60$ pF, $R_N  = 3.68 \Omega$, and $v I_C=0.0424$A/s.  
In this calculation we have assumed $R_s=R_N$ for simplicity.
Fig. 3 shows the dependence of $T^*$ on the mismatch angle $\chi$.
By changing $\chi$, we can control the macroscopic quantum dynamics, i.e., the MQT rate of the junction artificially.
Note that the maximum $T^*$ is attained when $\chi=0$.
Although $T^*$ is reduced by the dissipative effects, its magnitude  ($T^* \approx 0.12$K at $\chi=0$) is comparable to high quality $s$/$s$ junctions ($T^* \approx 0.3$K)~\cite{rf:Wallraff} and much larger than YBCO junctions ($T^* \approx 0.05$K).~\cite{rf:Bauch1}
This can be attributed to the high Josephson critical current density $j_C \approx 14$kA/cm${}^2$ and the weak Ohmic dissipative effect despite the small $\beta_M$.
Therefore, MQT should still be experimentally observable using the current junction fabrication and measurement technology.
Moreover, if we increase $j_C$ or decrease the capacitance per area $C/A$ of the junction, we can get still larger $T^*$.

\section{Summary}

In conclusion, MQT in the  $s$/$d$ hybrid Josephson junctions with the perfect insulating barrier has been theoretically investigated using the path integral method.
The effect of the low energy quasiparticles on MQT is found to be weak.
 This can be attributed to the quasiparticle-tunneling blockade effect in the $s$-wave superconductor.
 We also investigated MQT in a realistic $s$/$d$ junction and showed that the expected $T^*$ is relatively high in spite of the small $\beta_M$.
  These results strongly indicate the high promise of $s$/$d$ hybrid junctions for quantum computer applications.

   Finally we would like to comment the advantage of the $s/d$ junctions over the $s/s$ and $d/d$ junctions.
As compared to $s/s$ junction, $s/d$ junctions with the clean interface will have much larger $I_c$.
So we can expect that such $s/d$ junctions show higher $T^*$ than the $s/s$ junctions.
In $d_0/d_0$ junctions, the nodal-quasiparticles  give negligibly small effect on MQT.~\cite{rf:Kawabata1,rf:Yokoyama} 
 On the other hand, as was shown by Fominov $et.$ $al.$,~\cite{rf:Fominov} the decoherence time of the $d_0/d_0$ qubit is not enough for practical quantum computation.
 This indicates that the nodal quasiparticles still have large influence on the qubit operation.
However, due to the quasiparticle tunneling blockade effect in the $s/d$ junctions, the decoherence time of the $s/d$ qubit is expected to be much longer than that of the $d_0/d_0$ qubits.
Detailed discussion along this line is an interesting future problem.

\acknowledgments

  We would like to thank Y. Asano, T. Bauch, A. Brinkman, S. Kashiwaya, T. Kato, F. Lombardi, H. J. H. Smilde and Y. Tanaka for useful discussions. 
 One of the authors (S. K.) would like to express deep gratitude to S. Abe and H. Yokoyama for their encouragement.
  This work was supported by the Nano-NED Program under Project No. TCS.7029 and a Grant-in-Aid for Scientific Research from the Ministry of Education, Science, Sports and Culture of Japan (grant No. 17710081).

\end{document}